\documentclass[journal ]{new-aiaa}
\usepackage[utf8]{inputenc}
\usepackage{textcomp}
\usepackage{graphicx}
\usepackage{amsmath}
\usepackage[version=4]{mhchem}
\usepackage{siunitx}
\usepackage{longtable,tabularx}
\usepackage{multirow}
\usepackage{arydshln}
\usepackage{xcolor}

\definecolor{DarkGreen}{RGB}{83,129,53}

\usepackage{subfig}  

\title{Data-informed lifting line theory}
\author{%
  Arjun Sharma\footnote{Postdoctoral Appointee, Applied and Computational Mathematics Department. \textit{Corresponding author:} asharm1@sandia.gov}\,,\;
	Jonas A.\ Actor\footnote{Senior Member of Technical Staff, Scientific Machine Learning Department.}, and 
	Peter A.\ Bosler\footnote{Principal Member of Technical Staff, Applied and Computational Mathematics Department.}%
}

\affil{Center for Computing Research, Sandia National Laboratories, Albuquerque, NM, 87185, USA}

\begin{document}

\maketitle
\begingroup
\footnotesize
\noindent\textit{Accepted manuscript for publication in \emph{AIAA Journal of Aircraft}. This arXiv posting is the accepted manuscript and not the version of record.}

\medskip
\noindent Copyright © 2026 by the American Institute of Aeronautics and Astronautics, Inc. The U.S. Government has a royalty-free license to exercise all rights under the copyright claimed herein for Governmental purposes. All other rights are reserved by the copyright owner.

\endgroup

\begin{abstract}
We present a data-driven framework that extends the predictive capability of classical lifting-line theory (LLT) to a wider aerodynamic regime by incorporating higher-fidelity aerodynamic data from panel method simulations. A neural network architecture with a convolutional layer followed by fully connected layers is developed, comprising two parallel subnetworks to separately process spanwise collocation points and global geometric/aerodynamic inputs such as angle of attack, chord, twist, airfoil distribution, and sweep. Among several configurations tested, this architecture is most effective in learning corrections to LLT outputs. The trained model captures higher-order three-dimensional effects in spanwise lift and drag distributions in regimes where LLT is inaccurate, such as low aspect ratios and high sweep, and generalizes well to wing configurations outside both the LLT regime and the training data range. The method retains LLT’s computational efficiency, enabling integration into aerodynamic optimization loops and early-stage aircraft design studies. This approach offers a practical path for embedding high-fidelity corrections into low-order methods and may be extended to other aerodynamic prediction tasks, such as propeller performance.
\end{abstract}

{\section*{Nomenclature}
\begin{longtable*}{@{}l @{\quad=\quad} l@{}}
	$AR$                       & Aspect ratio, $b^2/S$ \\
	$b$                        & Wingspan (physical) \\
	$S$                        & Planform area \\
	$y$                        & Normalized spanwise coordinate ($-1 \le y \le 1$) \\
	$c(y)$                     & Chord length at $y$ \\
	$\Gamma(y)$                & Bound circulation at $y$ \\
	$l(y)$                     & Lift per unit span at $y$ \\
	$d(y)$                     & Induced drag per unit span at $y$ \\
	$L$                        & Total lift \\
	$D$                        & Total drag \\
	$V_\infty$                 & Freestream speed \\
	$\rho_\infty$              & Freestream density\\
	{$\bar{c}$} & {Average of chord length at wing root $(c_\text{root})$ and tip $(c_\text{tip})$}\\
	$c_l(y)$                   & Sectional lift coefficient at $y$ {based on $\bar{c}$} \\
	$c_d(y)$                   & Sectional induced–drag coefficient at $y$ {based on $\bar{c}$} \\
	{$\lambda$}				& {Taper ratio (root to tip chord)} \\
	$C_L$                      & Total lift coefficient \\
	$C_D$                      & Total drag coefficient \\
	$a_0(y)$                   & Sectional lift-curve slope at $y$ \\
	$\alpha_{L=0}(y)$          & Sectional zero-lift angle at $y$ \\
	$\alpha_\infty$            & Freestream angle of attack (relative to wing) \\
	$\alpha_{\mathrm{eff}}$    & Effective angle of attack at the section \\
	$\Lambda(y)$               & Sweep angle (section or quarter-chord), degrees unless noted \\
	$\theta(y)$                & Geometric twist distribution, degrees unless noted \\
	$N$                        & Number of spanwise collocation points (training used $N=40$) \\
	$N_{\text{out}}$           & Network output size ($2N{+}2$) \\
	$N_s$                      & Number of training/validation samples \\
	$x_i^{(1)}$                & Collocation-branch input vector (spanwise coordinates) \\
	$x_i^{(2)}$                & Wing\&\,flow-branch input (AoA, AR, sweep, chord, twist, $a_0$, $\alpha_{L=0}$) \\
	$f_{\theta}^{\mathrm{BB}}(x)$ & Black-box network output \\
	$f_{\theta}^{\mathrm{GB}}(x)$ & Grey-box network (LLT-residual) output \\
	$P$                        & Number of trainable parameters in the network \\
\end{longtable*}}

\section{Introduction}\label{sec:Intro}
In preliminary aircraft design, aerodynamic predictions must be obtained quickly to enable rapid iteration across a wide design space. Prandtl's lifting-line theory (LLT) \citep{PRANDTL1918,prandtl1923applications} has long been favored in this setting for its minimal computational cost and simple geometric inputs, yet it becomes increasingly inaccurate for wings with low aspect ratio, sweep, or complex twist. This paper introduces a machine-learning–based correction to LLT that preserves its simplicity while extending its accuracy to configurations that traditionally require higher-fidelity methods.

LLT is a classical reduced-order model for computing lift on a finite, three-dimensional wing in inviscid, incompressible flow. {It requires only a limited set of geometric parameters, such as chord and twist distributions.} {LLT can be modified into nonlinear LLT, in which two-dimensional (2D) section lift curves measured beyond the linear range are supplied to the lifting-line equation to incorporate post-stall behavior~\cite{sivells1947method}.} The original formulation was developed for straight wings and does not account for sweep, dihedral angle, or sideslip. {Phenomena beyond the linear regime, e.g., spanwise lift cells observed in the post-stall, are also outside classical LLT.} Over time, several extensions have been proposed to incorporate these factors more accurately {\cite{lan1973improved,lan1977applications,phillips2000modern,spalart2014prediction,reid2021general,goates2021practical}}. Despite its simplifying assumptions, LLT {(and its modified versions)} remains an invaluable tool for preliminary design, where computational efficiency is prioritized over the higher accuracy offered by panel methods or computational fluid dynamics (CFD). This trade-off is often acceptable during early-stage aerodynamic optimization, where the design space is large. LLT continues to be useful in applications such as turboprop wing-shape optimization \cite{rakshith2015optimal,sharma2024wing}, morphing-wing design \cite{snow2021design}, and flapping-wing efficiency studies \cite{bhowmik2013aerodynamic}. 

{Kontogiannis and Laurendeau~\cite{kontogiannis2021adjoint} use an adjoint nonlinear vortex-lattice method (NL-VLM), a lifting-surface generalization of lifting-line, to assimilate three dimensional RANS spanwise-load data by optimally adjusting geometric twist, thereby calibrating NL-VLM and improving agreement with RANS.} {Complementary to such adjoint-based approaches, when high-fidelity aerodynamic data (e.g., wind-tunnel tests, VLM, panel methods, or CFD) are available, machine-learning (ML) can enhance lower order methods such as LLT by learning data-informed corrections without implementing adjoint sensitivities.} This hybrid approach retains the simplicity of LLT while significantly improving its accuracy. In this study, we develop a physics-informed ML method to correct LLT predictions using data from the high-fidelity PANAIR panel code \citep{epton1990pan}. The resulting model preserves LLT’s input–output structure but approaches the predictive accuracy of PANAIR. {Our data-informed LLT parallels the historical nonlinear LLT in structure as both prepend a qualifying modifier to Prandtl’s base LLT to extend its regime of accuracy.}

As observed in trailing vortices behind airliners and in wind-tunnel experiments, a wing sheds vorticity from its trailing edge due to leakage of flow from the high-pressure lower surface to the low-pressure upper surface, especially near the wingtips \cite{anderson2017fundamentals}. Prandtl modeled this behavior by replacing the wing surface with a bound vortex distributed along the span, located at the aerodynamic center of the airfoil section (typically the quarter-chord point), and two trailing vortices extending downstream from the wingtips, thereby forming a horseshoe vortex. For an unswept wing, this bound vortex is straight. The local circulation $\Gamma(y)$ at spanwise location $y$ produces a sectional lift per unit span given by $L'(y) = \rho_\infty V_\infty \Gamma(y)$, where $V_\infty$ is the freestream velocity and $\rho_\infty$ is the fluid density. Assuming that each spanwise section behaves two-dimensionally, the lift can also be written as $L'(y) = \tfrac{1}{2} \rho_\infty V_\infty^2 c(y)\, c_l\!\left(y,\alpha_{\text{eff}}\right)$, where $c(y)$ is the local chord and $c_l$ is the airfoil lift coefficient at the effective angle of attack $\alpha_{\text{eff}}$.

Equating these expressions yields a precursor to the lifting-line equation. In the absence of trailing vortices, $\alpha_{\text{eff}}$ would be known from the flight angle of attack and the wing’s twist. However, by the Biot–Savart law, the trailing vortex sheet induces a vertical velocity (downwash), which reduces $\alpha_{\text{eff}}$ and thus the lift, while also tilting the lift vector rearward and producing induced drag. Incorporating this effect leads to the classical lifting-line equation \cite{anderson2017fundamentals}:
\begin{equation}
	\alpha_\infty = \frac{2\,\Gamma(y)}{a_0(y)\,V_\infty\,c(y)} + \alpha_{L=0}(y) + \frac{1}{2\,a_0(y)\,V_\infty} \int_{-b/2}^{b/2} \frac{d\Gamma/dy'}{y - y'}\,dy', \label{eq:ClassicLLTMain}
\end{equation}
where $a_0(y)$ is the lift-curve slope, $\alpha_{L=0}(y)$ is the zero-lift angle of attack of the local airfoil profile, and $\alpha_\infty$ is the freestream angle of attack relative to the wing. Nonlinear $c_l$–$\alpha$ relationships for the airfoil profile may also be incorporated numerically \citep{sivells1947method,lan1973improved}.

Perhaps the most important assumption of LLT is the requirement of a high aspect ratio (AR). {At low AR, the flow over each spanwise section is no longer quasi-two-dimensional and the theory becomes inaccurate.} \citet{snyder2005modified} introduced corrections for vortex roll-up near the wingtips, while \citet{hodson2019numerical} derived AR-dependent corrections to lift-curve slope for untwisted, unswept wings, showing good agreement with PANAIR. However, induced-drag predictions remain unreliable, highlighting the limitations of LLT even when corrected for lift-curve slope. From a perturbation-theory perspective, LLT can be seen as the leading-order solution in a singular perturbation expansion in inverse aspect ratio. Higher-order corrections using this approach have been developed by \citet{guermond1990generalized} and \citet{kida1978alternative}, yielding terms of order $1/\mathrm{AR}$ and $1/\mathrm{AR}\,\log(\mathrm{AR})$. These corrections improve agreement with higher-order lifting-surface methods even for wings with curved geometries and moderate AR values around 7.

Swept wings also exhibit additional three-dimensional features that were incorporated numerically by \citet{phillips2000modern} using a curved lifting line aligned with aerodynamic centers. However, this extension poses significant challenges. The geometric curvature of the lifting line and misalignment of trailing vortices lead to poor convergence in numerical solutions, as noted by \citet{reid2021general}. Furthermore, the Biot–Savart law predicts an infinite self-induced downwash when the bound vortex is curved \citep{mclean2012understanding}, creating further difficulties. Common fixes, such as displacing the control point to the three-quarter-chord location \citep{weissinger1947lift}, are only effective for thin, mildly cambered airfoils. To address the singularities (and hence convergence issues) that arise in the integral equations of LLT, \citet{reid2021general} introduced a local linearization technique that blends the lifting-line geometry and modifies the trailing-vortex orientation to remain perpendicular to the bound vortex. This method was further refined by \citet{goates2023modern}, who incorporated thin-airfoil theory to estimate effective section properties normal to the curved lifting line, thereby improving the applicability of LLT to swept-wing configurations. In this study, we adopt the extended LLT formulation developed by \citet{reid2021general} and \citet{goates2021practical} as the baseline. The data-driven corrections proposed here are applied on top of this improved version of LLT.

Inspired by recent work in physics-informed ML, we aim to improve LLT by blending it with high-fidelity panel-method data using neural networks. While deep learning has been widely used to approximate solutions of partial differential equations \citep{cai2021physics,wang2021learning}, the generalizability of such models across asymptotic regimes remains limited. A more effective approach is to formulate the ML task in a way that respects classical asymptotic structure. For example, \citet{martin2023physics} augmented the Kuramoto–Sivashinsky (KS) equation, a reduced-order model for thin-film flows derived via regular perturbation in film thickness, using data from two-dimensional Navier–Stokes simulations. Their ML-enhanced model preserved the form of the KS equation while significantly extending its predictive range beyond the asymptotic regime. Analogously, we use high-fidelity data from the PANAIR panel method to learn corrections to Prandtl’s LLT, which is itself a singular-perturbation model in inverse aspect ratio. This asymptotics-informed perspective guides our network design: we preserve the spanwise structure and physical inputs of LLT (chord, twist, sweep, airfoil parameters) while training the model to learn three-dimensional corrections. Our neural network takes LLT-style inputs and predicts corrected spanwise lift and drag distributions. Although data generation and training incur a one-time offline cost, the resulting ML-augmented LLT model generalizes well to previously unseen wing geometries and achieves significantly improved accuracy at a computational cost comparable to classical LLT. This makes it particularly suitable for preliminary design and optimization workflows.

A related effort is VortexNet \cite{shen2025vortexnet}, which uses graph neural networks to predict surface pressure-coefficient distributions over triangulated 3D lifting surfaces. While powerful, that approach requires mesh-based input and produces high-dimensional surface-level outputs. In contrast, our model operates on a one-dimensional lifting line and is trained in a grey-box setting (section~\ref{sec:method}) to learn corrections to LLT. This formulation improves interpretability, computational speed, and compatibility with existing LLT-based analysis and design tools. It also enables generalization across a wide range of planform geometries and airfoil families. ML-based aerodynamic modeling has also been applied to shock prediction \cite{sabater2022fast}, airfoil characterization \cite{zuo2023fast}, shape optimization \cite{li2022machine}, stall detection \cite{saetta2022machine}, and turbulence modeling \cite{li2025field}.

The rest of the paper is organized as follows. The neural-network design and methodology are discussed in section~\ref{sec:method}, and the dataset and criteria for model selection are described in section~\ref{sec:Data}. Section~\ref{sec:Results} illustrates the performance of the best-performing model, before conclusions and future directions are outlined in section~\ref{sec:Conclusions}.

\section{Methodology}\label{sec:method}
We develop a neural network model in PyTorch \citep{paszke2019pytorch} to construct a data-driven enhancement of lifting-line theory (LLT). The workflow begins by generating aerodynamic data over a wide range of wing geometries and flow conditions using PyPan \citep{pypan}, a Python-based implementation of the PANAIR panel method \citep{epton1990pan}. The corresponding LLT predictions are obtained using MachUpX \citep{machupx2025}, which implements an extended version of LLT that incorporates recent improvements \citep{reid2021general,goates2023modern}.

We consider two modeling strategies based on the training data: black-box and grey-box. In the black-box approach, the model is trained solely on PANAIR outputs. In the grey-box approach, both PANAIR and LLT outputs are incorporated, as described below. After assembling the datasets, we train and evaluate multiple neural network architectures to identify the best-performing configuration, which is discussed in the next section.

\subsection{Inputs}

\begin{figure*}[ht]
	\centering
	\includegraphics[width=0.75\linewidth]{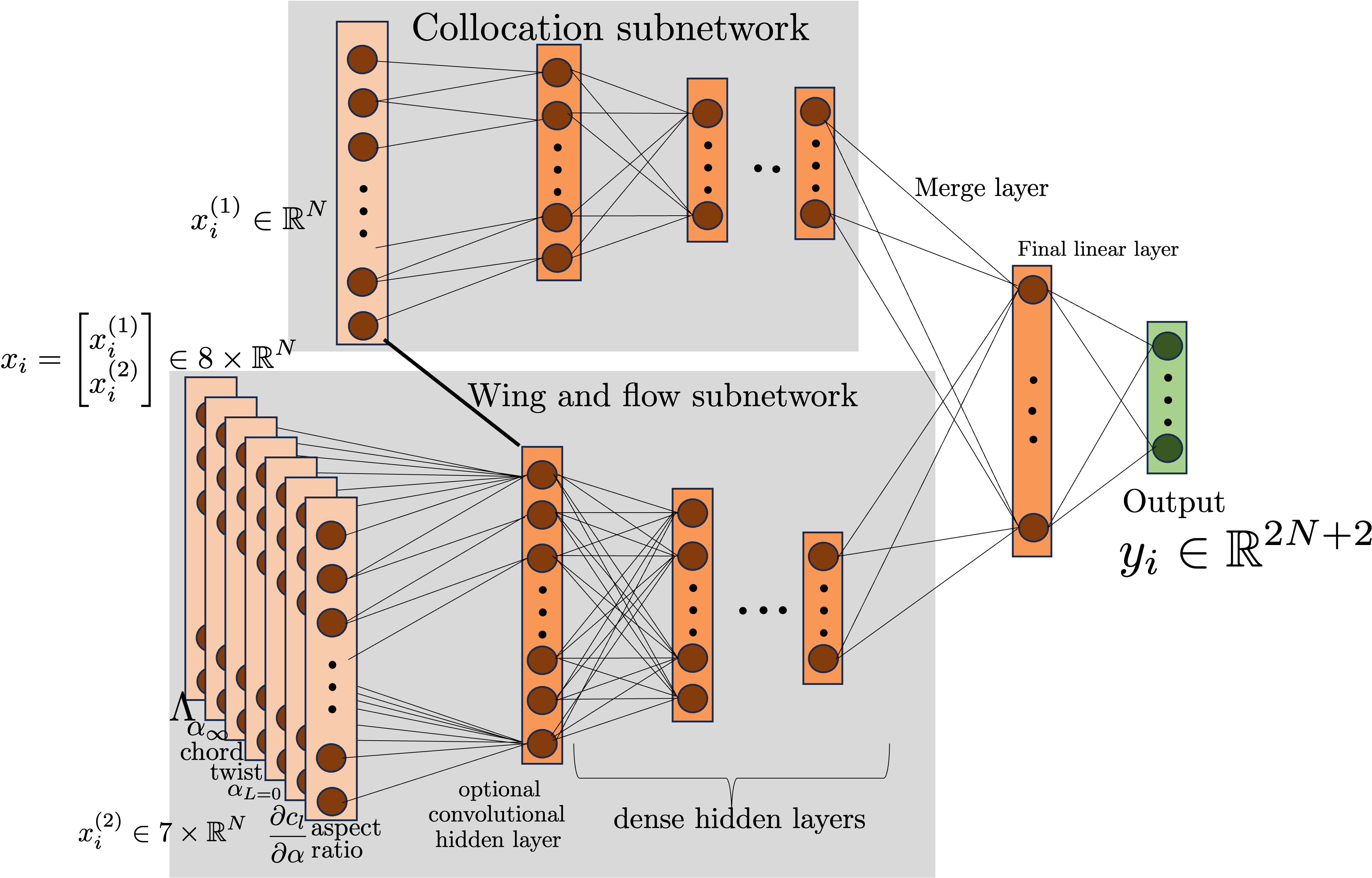}
	\caption{{Neural network architecture: two parallel subnetworks (collocation; wing and flow) combine after optional convolution/dense layers to output spanwise lift, spanwise drag, and totals.}}
	\label{fig:network}
\end{figure*}
{Fig.}~\ref{fig:network} shows a schematic of the neural network architecture, which consists of two parallel subnetworks that are merged before the final output layers. We use non-dimensional sectional lift and drag coefficients at each spanwise location,
\begin{equation}
	{c_l(y)=\frac{l(y)}{0.5 \rho_\infty V_\infty^2 \bar{c}},\hspace{0.2in} c_d(y)=\frac{d(y)}{0.5 \rho_\infty V_\infty^2  \bar{c}},}
\end{equation}
as well as non-dimensional total lift and drag coefficients,
\begin{equation}
	C_L=\frac{L}{0.5  \rho_\infty V_\infty^2 S}, \hspace{0.2in} C_D=\frac{D}{0.5  \rho_\infty V_\infty^2 S},
\end{equation}
where $l(y)$ and $d(y)$ are the spanwise lift and drag forces, {$\bar{c}=0.5(c_\text{root}+c_\text{tip})$ is the reference chord length defined as the average of root and tip chord,} and $S$ is the area of the wing planform. {For trapezoidal wings, $S=b \bar{c}$.}  We normalize the span to length 2. In our calculations, we set $\rho_\infty = 1$ and $V_\infty = 1$. The total lift $L$ and drag $D$ are obtained as
\begin{equation}
	L=\int_{-1}^1 l(y')\, dy', \hspace{0.2in} D=\int_{-1}^1 d(y')\, dy'.
\end{equation}

The collocation subnetwork processes the spanwise collocation points $y$, which correspond to the bound vortex locations in LLT. The input to this subnetwork is a vector $x_i^{(1)}$ of size $N \times 1$, with $N = 40$ matching the grid resolution used in training. The second branch, called the wing-and-flow subnetwork, accepts seven input channels representing geometric and aerodynamic properties: angle of attack ($\alpha_\infty$), aspect ratio, and spanwise distributions of sweep angle ($\Lambda$), chord, twist, sectional lift-curve slope ($\partial c_l / \partial \alpha$), and zero-lift angle ($\alpha_{L=0}$). This input is denoted $x_i^{(2)} \in 7 \times \mathbb{R}^{N}$ in {Fig.}~\ref{fig:network}. Although $\alpha_\infty$ and aspect ratio are scalar parameters, they are broadcast to vectors to match PyTorch input conventions. {Our training data is constrained to the linear part of the sectional lift curve, allowing us to parameterize the airfoil properties using just the $\partial c_l / \partial \alpha$ and $\alpha_{L=0}$. Thus, the trained model presented in the next section is valid away from the stall regime.}

Freestream velocity is not included as an input, since the data is non-dimensionalized. Future extensions may include spanwise variations in freestream velocity, such as those induced by upstream propellers \citep{rakshith2015optimal}, or additional geometric features like dihedral angle or nonlinear airfoil behavior \citep{moorthamers2019aerodynamic}, by introducing extra input channels.

\subsection{Neural network architecture}
As noted in the introduction, classical LLT assumes that sectional lift depends only on local flow conditions. This assumption breaks down for wings with low aspect ratio or large sweep, where spanwise crossflow and three-dimensional effects introduce non-local influences on the lift distribution. To account for such non-local interactions, we introduce cross-subnetwork coupling by concatenating the collocation input with the wing-and-flow subnetwork before the first hidden layer of the wing-and-flow subnetwork.

We evaluate four architectural variants: two with and two without an initial convolutional layer, each tested at both shallow and deep network depths. The convolutional layer is expected to improve learning of local patterns and is particularly relevant for high aspect ratio or unswept wings where non-local interactions are less prominent. Beyond the first layer, all architectures consist of fully connected layers. We use the \texttt{SiLU} (Sigmoid Linear Unit) activation function, which outperforms ReLU in our early tests (not shown). During training, the dataset is divided into mini-batches of 512 samples out of the total $N_s$ data samples. Here $N_s=4\times 10^5$ as described in section \ref{sec:Data}. This batching strategy allows each gradient update to be computed on a representative subset of the data, rather than the entire training set, enabling faster and more memory-efficient training. In addition, mini-batch updates introduce stochasticity into the optimization process, which can help escape shallow local minima and improve generalization performance. The loss expressions in upcoming {Eqs.~\ref{eq:BlackBoxLoss} and \ref{eq:GreyBoxLoss}} represent averages over a full batch, and are thus approximated during training using these mini-batch samples. To improve training stability and convergence, we apply Layer Normalization (using the PyTorch's \texttt{LayerNorm} function) after each hidden layer. LayerNorm normalizes the pre-activation outputs of each neuron across the feature dimension within a single training sample. Given an input vector $\mathbf{h} \in \mathbb{R}^d$, LayerNorm computes
	\begin{equation}
	\text{LayerNorm}(\mathbf{h}) = \frac{\mathbf{h} - \mu}{\sigma} \odot \boldsymbol{\gamma} + \mathbf{b},
	\end{equation}
	where $\mu$ and $\sigma$ are the mean and standard deviation of $\mathbf{h}$, and $\boldsymbol{\gamma}$ and $\mathbf{b}$ are learned parameters; $\odot$ denotes the elementwise multiplication \citep{ba2016layer}. Unlike Batch Normalization, which normalizes across the batch dimension and is sensitive to batch size, LayerNorm is independent of batch size and better suited for problems with heterogeneous inputs such as in our case, where input configurations may span a broad distribution of wing geometries and flow conditions. In our training, it improved convergence and stability (not shown). Weight and bias initialization use the Xavier-uniform and uniform distributions, respectively.

The deeper version of the network includes:
\begin{itemize}
	\item Collocation subnetwork: 4 hidden layers of sizes 256, 128, 64, and $N_{\text{out}}$
	\item Wing-and-flow subnetwork: 5 hidden layers of sizes 256, 128, 64, 64, and $N_{\text{out}}$
\end{itemize}

The shallower version includes:
\begin{itemize}
	\item Collocation subnetwork: 2 hidden layers of sizes 128 and $N_{\text{out}}$
	\item Wing-and-flow subnetwork: 3 hidden layers of sizes 256, 64, and $N_{\text{out}}$
\end{itemize}

For all configurations, the final output size of each subnetwork is $N_{\text{out}} = 2N + 2$ (spanwise $c_l$, spanwise $c_d$, and scalars $C_L$, $C_D$). These outputs are concatenated into a merged layer of size $2N_{\text{out}}$, which is followed by a final output layer of size $N_{\text{out}}$.

\subsection{Loss functions and outputs in black-box vs grey-box modeling}
The loss function is the mean squared error (MSE) between the network predictions and reference data. We consider two training paradigms: black-box and grey-box.

\paragraph{Black-box:}

In the black-box setting, the model output $f^{\text{BB}}_{\theta^{\text{BB}}} (x_i)$ consists of the normalized spanwise distributions of lift and drag computed by PANAIR. A normalization factor corresponding to the maximum spanwise lift or drag is also included. The network is trained by optimizing {$P$ parameters $\theta^{\text{BB}}$ over $N_s$} training samples:
\begin{equation}
	{\mathcal{L}_\text{BB}(x,y_i^\text{H},\theta^\text{BB})=\min_{\theta^{\text{BB}}\in\mathbb{R}^P}  \frac{1}{N_s} \sum_{i=1}^{N_s} \left\| f^{\text{BB}}_{\theta^{\text{BB}}} (x_i)- y_i^{\text{H}} \right\|_2^2,}\label{eq:BlackBoxLoss}
\end{equation}
where $y_i^{\text{H}}$ is the high-fidelity reference from PANAIR and $f^{\text{BB}}_{\theta^{\text{BB}}} (x_i)$ is the network prediction. 

\paragraph{Grey-box:}

{In the grey-box setting, the network learns to predict the difference between the higher-fidelity PANAIR output and the lower-fidelity LLT prediction. While the architecture may mirror that of the black-box model, the learning objective is different. Since this difference tends to vanish at high aspect ratios, where LLT becomes more accurate, we scale the normalization factor by the aspect ratio to avoid vanishing learning targets and improve numerical stability. The grey-box network is trained by minimizing}
\begin{equation}
	{\mathcal{L}_\text{GB}(x,y_i^\text{L},y_i^\text{H},\theta^\text{GB})=\min_{\theta^{\text{GB}}\in\mathbb{R}^P}  \frac{1}{N_s} \sum_{i=1}^{N_s} \left\| f^{\text{GB}}_{\theta^{\text{GB}}} (x_i)- \left(y_i^{\text{H}} - y_i^{\text{L}}\right) \right\|_2^2,}\label{eq:GreyBoxLoss}
\end{equation}
where $y_i^{\text{L}}$ and $y_i^{\text{H}}$ are the normalized outputs from LLT and PANAIR, respectively, including the AR scaling. The superscripts L and H represent the low and high fidelity model. Since the model explicitly incorporates LLT information to learn the correction, we refer to it as a grey-box model—intermediate in interpretability between a fully transparent physics-based model and an opaque black-box model.

\subsection{Training setup and hyperparameters (learning rate)}

Training uses double-precision tensors on a GPU, with a batch size of 512. We use the Adam optimizer with an initial learning rate of 0.001, 0.0001, or 0.000025. A \texttt{ReduceLROnPlateau} scheduler lowers the learning rate by a factor of 0.5 if the validation loss does not improve for 10 consecutive epochs, down to a minimum of one-twentieth of the initial value.
 
 \subsection{ {On choosing neural networks over multidimensional polynomials}}
 {A global multivariate polynomial is an appealing alternative to neural networks, but simple scaling shows it is infeasible for our field-to-field map. With $d=203$ input features (40 collocation points $y$; four spanwise fields (chord, twist, $\partial c_\ell/\partial\alpha$, and $\alpha_{L=0}$) each sampled at 40 points; and three global scalars (sweep, AR, and $\alpha_\infty$)), a total-degree-$k$ polynomial requires $\binom{d+k}{k}$ coefficients per output. Even $k=2$ requires $\binom{205}{2}=20{,}910$ coefficients per output ($\approx1.71\times10^6$ across our 82 outputs), and $k=3$ needs $\binom{206}{3}=1{,}435{,}820$ per output ($\approx1.18\times10^8$ total). Solving such least-squares systems on $400{,}000$ samples is prohibitive; per-geometry fits would also forfeit the advantage of a single reusable surrogate. By contrast, our largest neural network (deeper, with a convolutional layer) has 356{,}651 trainable parameters and the smallest (shallower, without convolution) has 128{,}635. Convolutional weight sharing encodes spanwise locality with far fewer degrees of freedom, and mini-batch optimization trains efficiently on large datasets. The result is a compact, reusable corrector with a fixed-cost forward pass.}

\section{Dataset and model selection}\label{sec:Data}
We divide the dataset into \textit{in-distribution} and \textit{out-of-distribution} samples. Training and validation are performed exclusively on the in-distribution set, while the test set includes a subset of the in-distribution samples and all out-of-distribution cases. This separation allows us to evaluate model performance on both familiar and novel samples. In this section, we describe the training and validation procedure used to select the optimal network architecture.

The \textit{in-distribution} dataset comprises 1,056,000 unique wing–flow configurations obtained by Cartesian product of all parameters listed in table~\ref{tab:param_ranges}. {From these, we preselect $N_s=400{,}000$ cases for model development, split 80/20 into 320{,}000 training samples (used to update the neural-network weights) and 80{,}000 validation samples (used to monitor validation loss and adjust the learning rate via the scheduler).} {The remaining 656{,}000 in-distribution cases are held out for testing and are never used during training or validation.} This ensures a rigorous assessment of generalization in table~\ref{tab:TestLossComparison}. 

\begin{table*}[ht]
	\centering
	\caption{Parameters used to generate the in-distribution dataset. A random subset of 400,000 samples is used for training and validation; the remainder are included in the test set.}
	\label{tab:param_ranges}
	\begin{tabular}{lc}
		\hline
		\textbf{Parameter} & \textbf{Values}  \\
		\hline
		{AR(1+$\lambda$)/2}      & 4, 5, 7.5, 10, 15, 20, 30, 40, 50, 75, 100  \\
		Angle of Attack (AoA)  & $-6^\circ$ to $6^\circ$ (step $0.05^\circ$)  \\
		Taper Ratio  ($\lambda$)          & 0.25, 0.5, 0.75, 1.0                      \\
		Sweep Angle            & $-30^\circ$, $-15^\circ$, $0^\circ$, $15^\circ$, $30^\circ$  \\
		Twist                  & $-5^\circ$, $-2.5^\circ$, $0^\circ$, $2.5^\circ$, $5^\circ$  \\
		Airfoils               & NACA 2440, 0040, 0020, 1320                  \\
		\hline
	\end{tabular}
\end{table*}

\begin{figure*}[ht]
	\centering
	\subfloat{\includegraphics[width=0.48\textwidth]{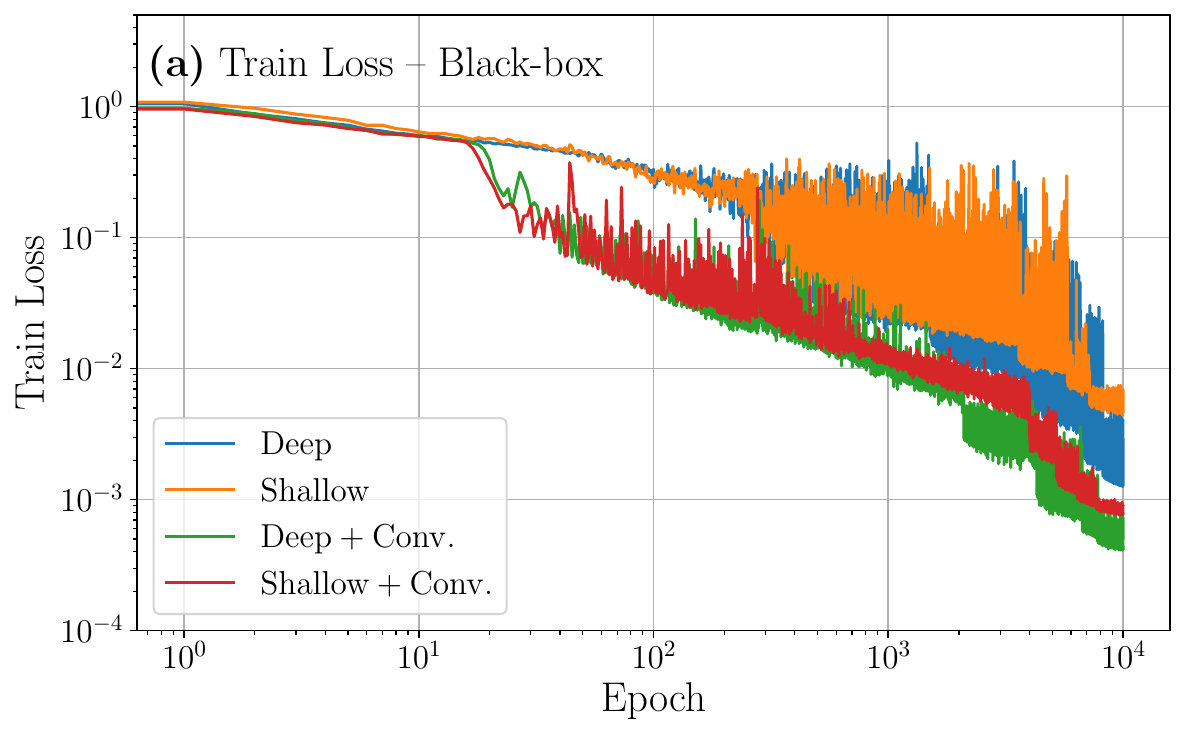}}
	\hfill
	\subfloat{\includegraphics[width=0.48\textwidth]{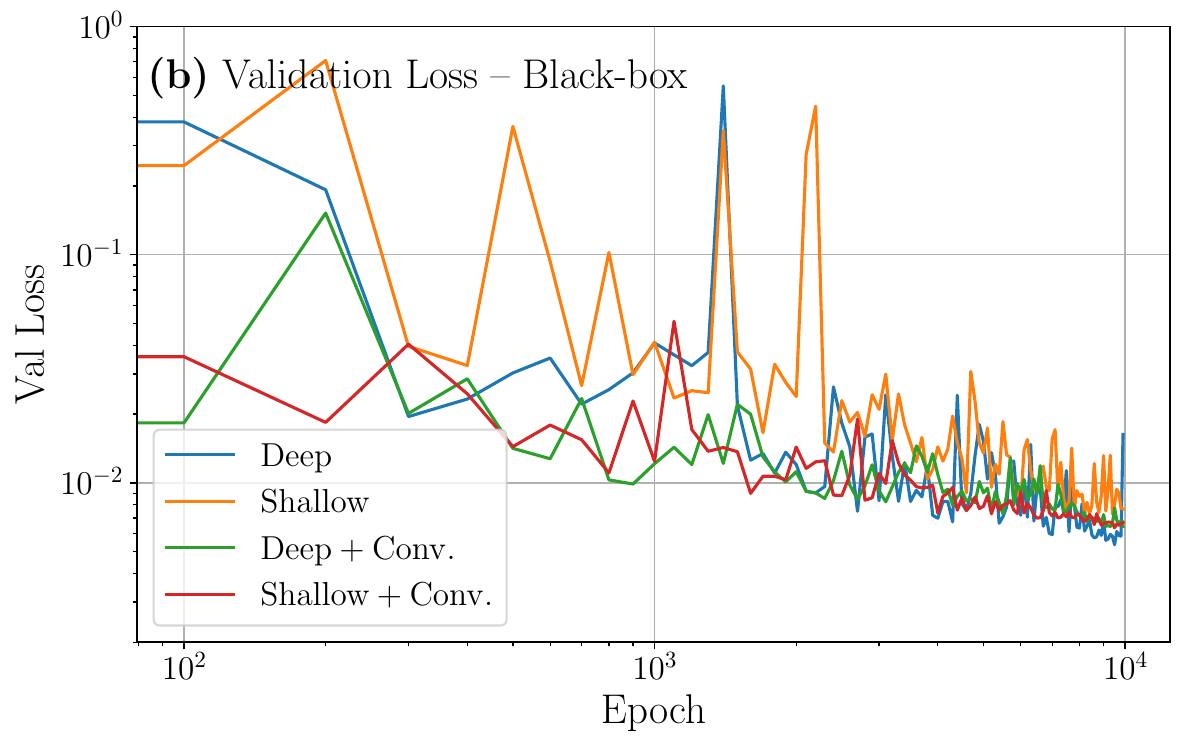}}	
	\vspace{0.5em}	
	\subfloat{\includegraphics[width=0.48\textwidth]{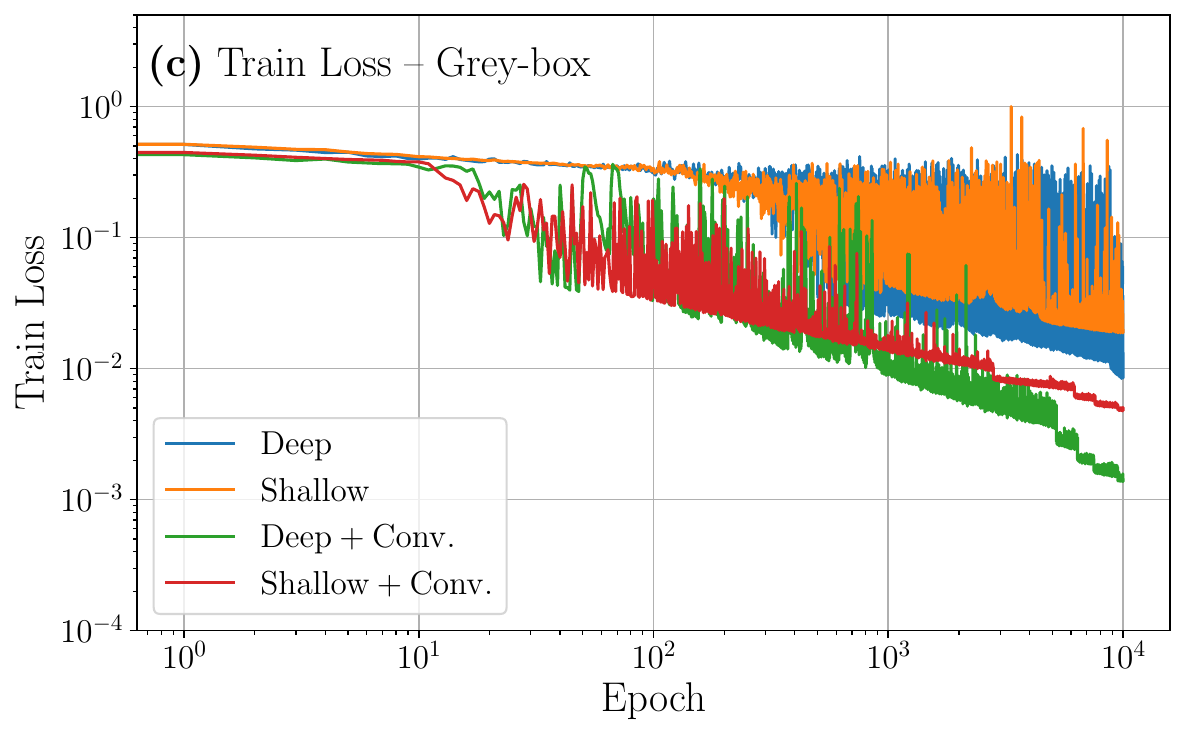}}
	\hfill
	\subfloat{\includegraphics[width=0.48\textwidth]{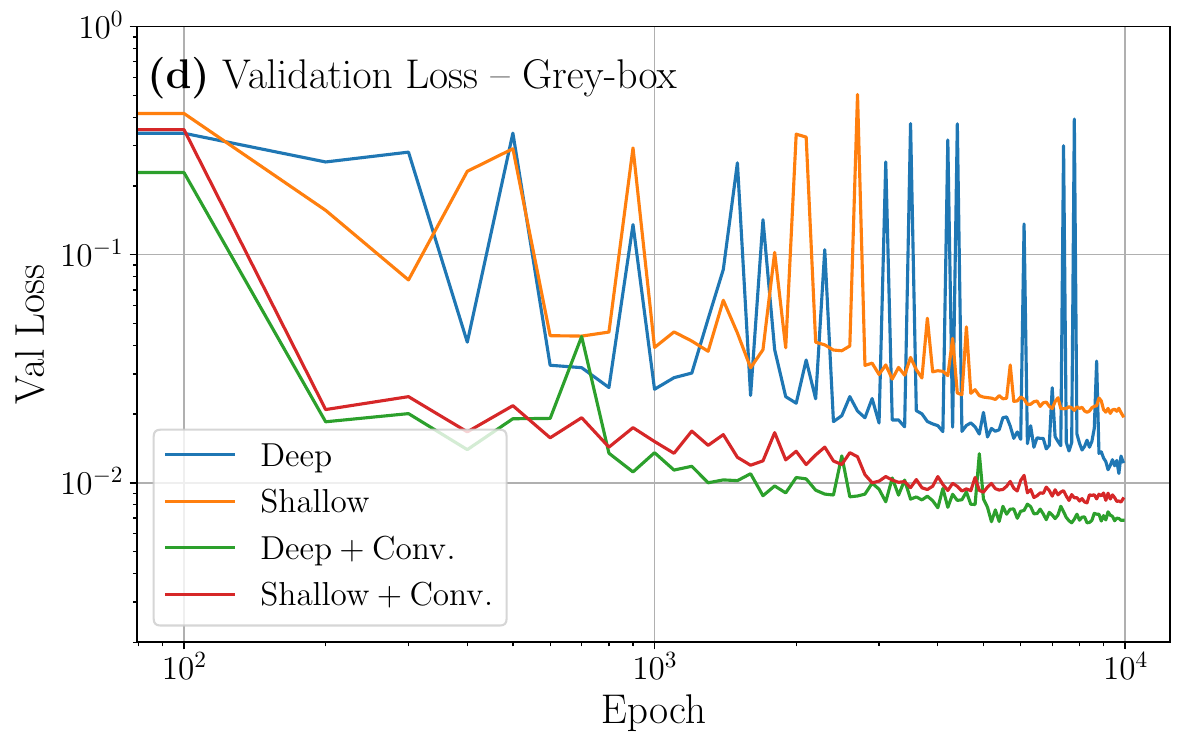}}	
	\caption{Training (a,c) and validation (b,d) loss histories for four architectures under black-box and grey-box formulations with an initial learning rate of 0.001.}
	\label{fig:loss_histories}
\end{figure*}

Validation data is used only for monitoring convergence and scheduling learning rate reductions; it is not used during gradient updates, thereby preserving its role in assessing generalization during training. Among the learning rates tested (0.001, 0.0001, 0.000025), the two highest rates yielded similar convergence, although 0.001 exhibits slightly better generalization. {Fig.}~\ref{fig:loss_histories} shows the training and validation loss histories for a learning rate of 0.001. For both black-box and grey-box loss formulations, architectures that include a convolutional layer (architectures 3 and 4) converge faster and achieve lower losses than their fully connected counterparts. In training, black-box models achieve lower loss than grey-box models (compare  {Figs.}~\ref{fig:loss_histories}a and \ref{fig:loss_histories}c), but in validation, architectures 3 and 4 perform similarly across both formulations (compare {Figs.}~\ref{fig:loss_histories}b and \ref{fig:loss_histories}d).

\begin{table*}[ht]
	\centering
	\caption{Test loss and relative error across network architectures and modeling settings.}
	\label{tab:TestLossComparison}
	\begin{tabular}{c c c c c c c }
		\hline\hline
		\textbf{Architecture} &	\textbf{Conv Layer} & \textbf{Depth} & \textbf{Model Type} & 
		\textbf{test loss} & 
		\textbf{Rel. Err.}& 
		\textbf{Rel. Err. GB/BB}  \\
		\hline
		1  & Absent  & Deep     & Black-box & 8.261e-03 & 2.306e-01& \multirow{2}{*}{0.54}  \\
		1  & Absent  & Deep     & Grey-box  & 1.106e-02 & 1.239e-01&  \\
		\hdashline
		2  & Absent  & Shallow  & Black-box & 7.281e-03 & 3.163e-01& \multirow{2}{*}{0.57}  \\
		2  & Absent  & Shallow  & Grey-box  & 2.019e-02 & 1.794e-01&  \\
		\hdashline
		3  & Present & Deep     & Black-box & 7.262e-03 & 1.446e-01& \multirow{2}{*}{0.42}  \\
		3  & Present & Deep     & Grey-box  & \textbf{6.632e-03} & \textbf{6.076e-02 }& \\
		\hdashline
		4  & Present & Shallow  & Black-box & 6.859e-03 & 1.884e-01& \multirow{2}{*}{0.52}  \\
		4  & Present & Shallow  & Grey-box  & 8.540e-03 & 9.851e-02&  \\
		\hline\hline
	\end{tabular}
\end{table*}

{Table~\ref{tab:TestLossComparison} reports the test loss and relative error for each architecture and modeling type, selected based on the lowest validation loss}. The relative error is defined as
\begin{equation}
{	\text{Relative Error} = \frac{1}{2} \left( 
	\frac{||C_L^{\text{NN}} - C_L^{\text{H}}||_2}{||C_L^{\text{L}} - C_L^{\text{H}}||_2} + 
	\frac{||C_D^{\text{NN}} - C_D^{\text{H}}||_2}{||C_D^{\text{L}} - C_D^{\text{H}}||_2}
	\right),}
\end{equation}
where the norm is computed across the test set. {Here $C_L^{\text{L}}$ and $C_D^{\text{L}}$ represent the lift and drag coefficients from the low fidelity model, LLT, and $C_L^{\text{H}}$ and $C_D^{\text{H}}$ represent these coefficients from a higher fidelity PANAIR.} A relative error close to 1 indicates no improvement over LLT, while a value near 0 reflects strong agreement with PANAIR. For example, if the test set included only high aspect ratio, unswept wings, where LLT is already accurate, the relative error would be expected to approach 1.

Although validation losses are similar for architectures 3 and 4 across both training paradigms, the test metrics clearly favor the grey-box approach {as the ratio of the relative error of grey training for each of the four architectures is about 50\% lower than its black box counterpart.} Architecture 3 with grey-box training achieves the lowest test loss and relative error, and is therefore selected as the best-performing model. In the next section, we evaluate this model’s predictive performance across a range of unseen configurations. In addition to in-distribution cases, we assess extrapolation to out-of-distribution conditions involving sweep angle of $45^\circ$, twist angles of $\pm 7.5^\circ$, taper ratios of 0.1, and aspect ratios between 1.5 and 3.5.

\section{Results: Aerodynamic prediction from the best-performing network}\label{sec:Results}
\begin{figure*}[ht]
\centering	
\subfloat{\includegraphics[width=0.49\textwidth]{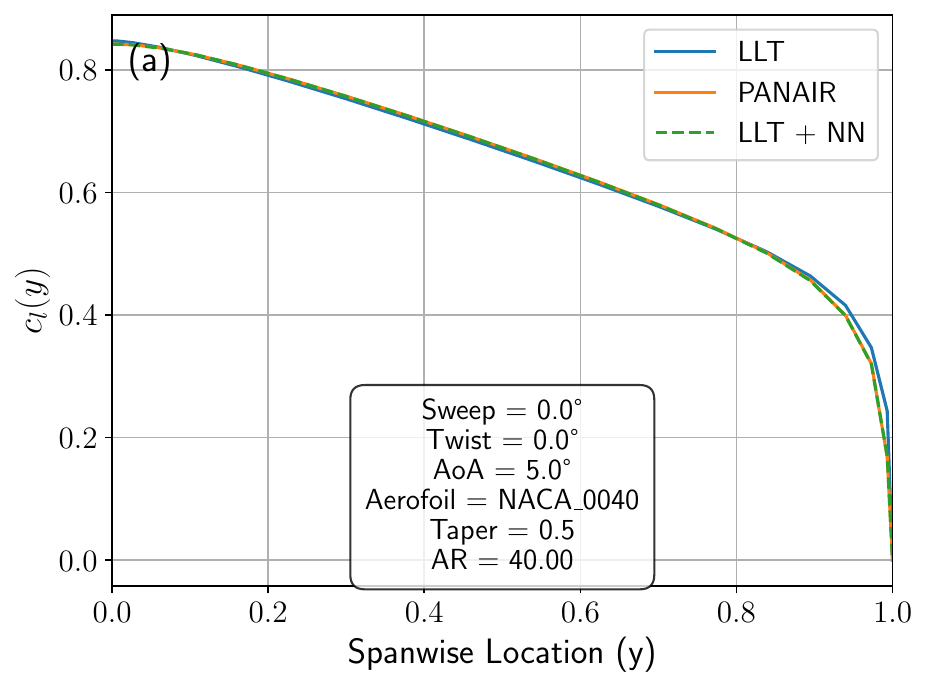}\label{fig:SpanwiseCLa}}\hfill
\subfloat{\includegraphics[width=0.49\textwidth]{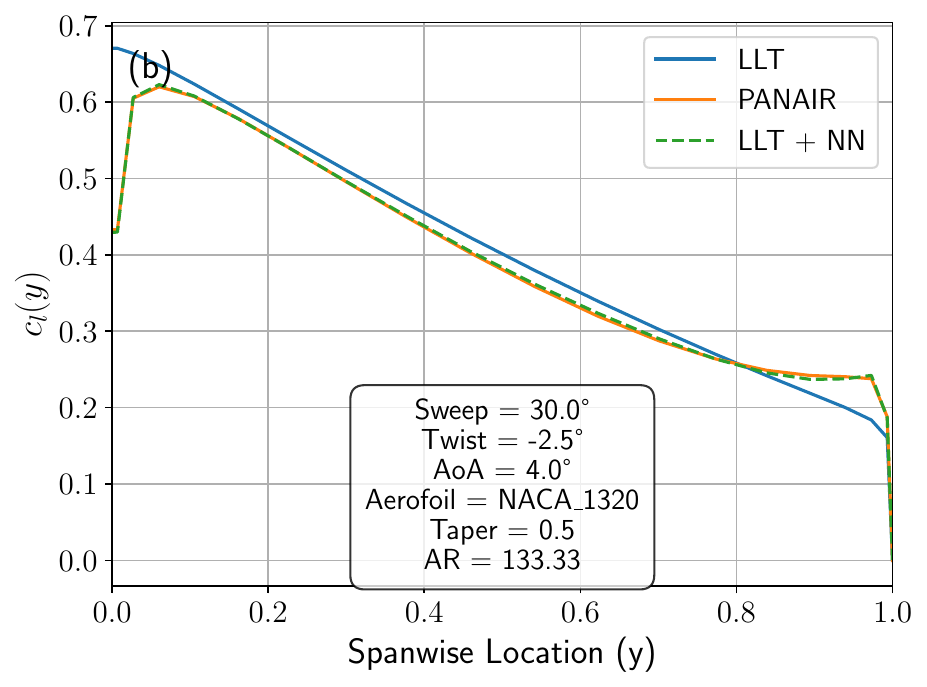}\label{fig:SpanwiseCLb}}\hfill
\subfloat{\includegraphics[width=0.49\textwidth]{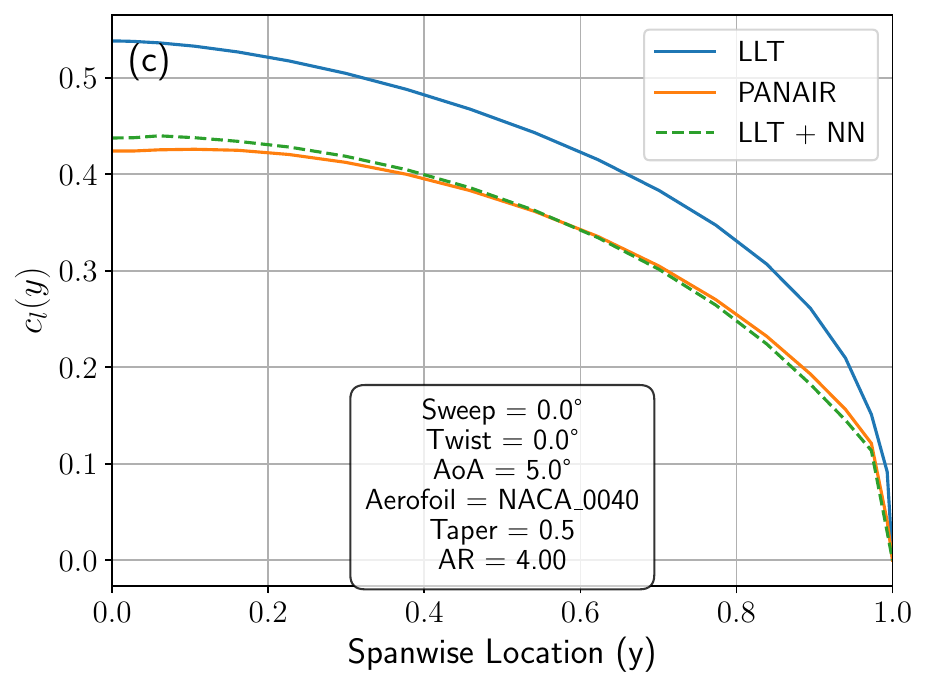}\label{fig:SpanwiseCLc}}\hfill
\subfloat{\includegraphics[width=0.49\textwidth]{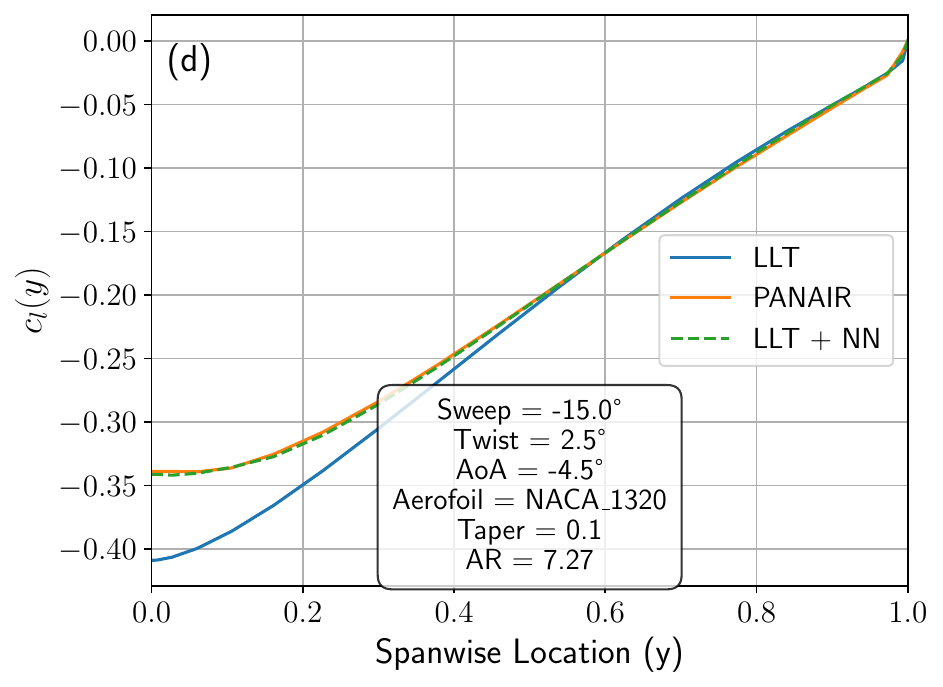}\label{fig:SpanwiseCLd}}\hfill
\subfloat{\includegraphics[width=0.49\textwidth]{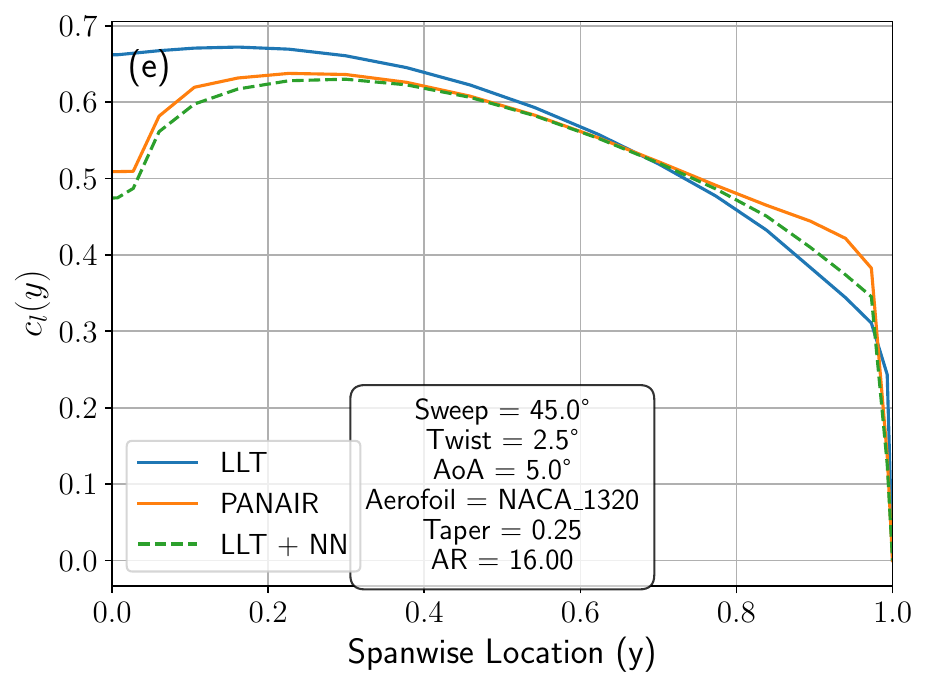}\label{fig:SpanwiseCLe}}\hfill
\subfloat{\includegraphics[width=0.49\textwidth]{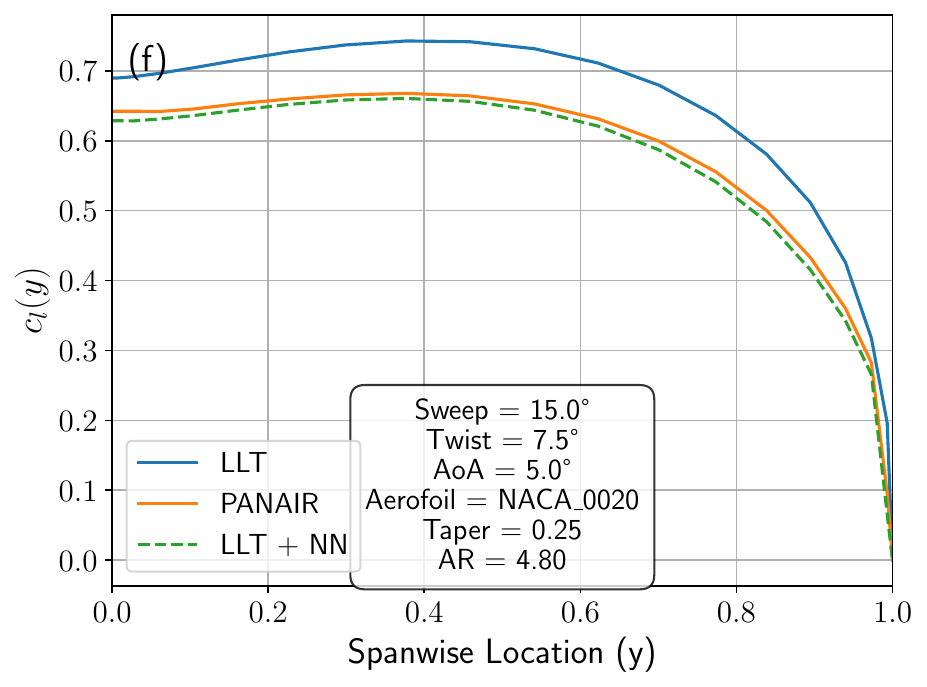}\label{fig:SpanwiseCLf}}
\caption{Spanwise lift coefficient $c_l(y)$ distributions for different wing and flow configurations labeled on each panel.}
\label{fig:SpanwiseCL}
\end{figure*}

\begin{figure*}[htbp]
	\centering	
\subfloat{\includegraphics[width=0.49\textwidth]{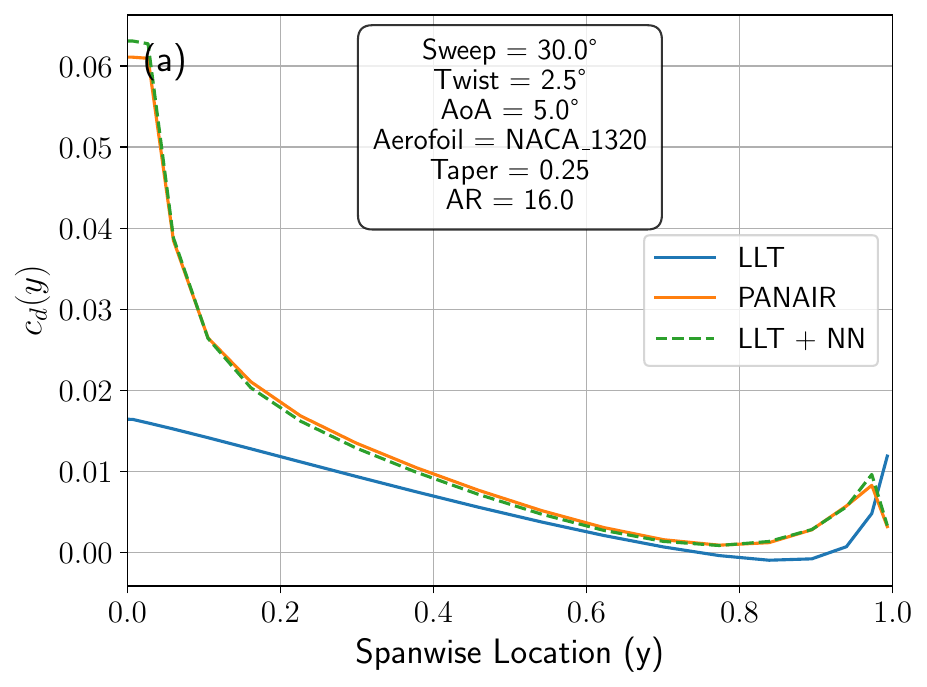}}\hfill
\subfloat{\includegraphics[width=0.49\textwidth]{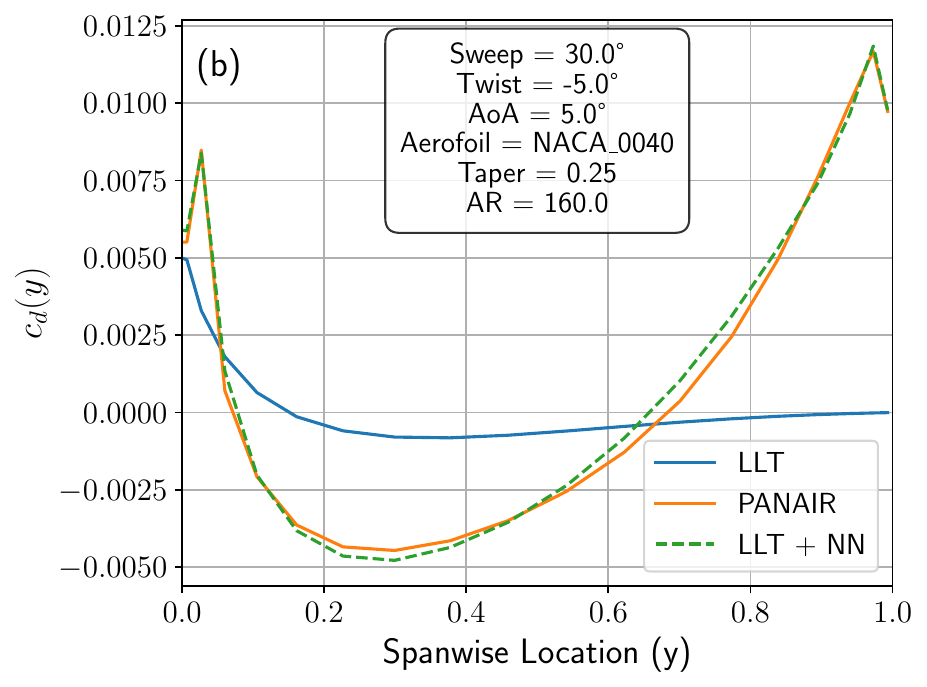}}\hfill
\subfloat{\includegraphics[width=0.49\textwidth]{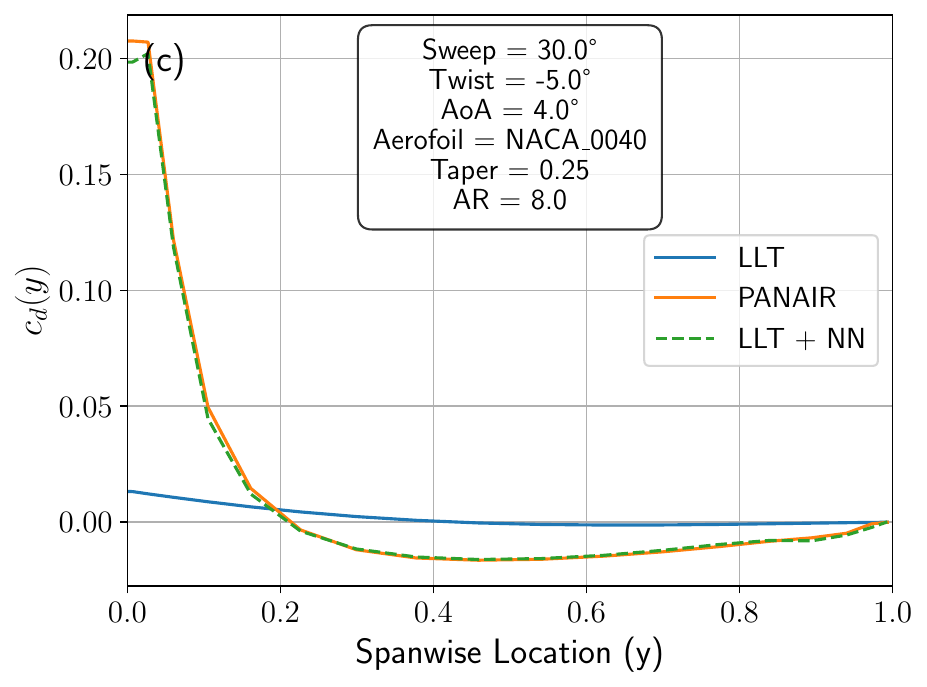}}\hfill
\caption{Spanwise induced-drag coefficient $c_d(y)$ distributions for different wing and flow configurations labeled on each panel.}
\label{fig:SpanwiseCD}
\end{figure*}

Spanwise lift and drag distributions for representative wing configurations are shown in Figs.~\ref{fig:SpanwiseCL} and \ref{fig:SpanwiseCD}. These plots compare predictions from lifting-line theory (LLT) \citep{reid2021general,goates2023modern}, the PANAIR panel method \citep{pypan,epton1990pan}, and the grey-box neural network corresponding to Architecture~3 discussed previously. We refer to these as LLT, PANAIR, and LLT+NN, respectively, where LLT+NN denotes that the network learns residual corrections to LLT outputs. Across the cases detailed below, LLT+NN closely tracks PANAIR, including configurations outside the training and validation ranges.

Since we do not consider sideslip or yaw in these examples, the spanwise force distributions are symmetric about the centerline of the wing planform, and we show only the right semi-span. We begin with a case where LLT (including the classical Prandtl formulation) is highly accurate. {Fig.}~\ref{fig:SpanwiseCLa} shows an unswept wing with aspect ratio (AR) 30, for which all three methods produce nearly identical lift distributions. This demonstrates that when no correction is required, the grey-box network yields effectively zero residual. In contrast, at AR {$=4.0$}, which lies well outside LLT’s formal validity (Fig.~\ref{fig:SpanwiseCLc}), PANAIR predicts reduced lift due to stronger three-dimensional effects; LLT+NN reproduces this trend.

As described in section~\ref{sec:Intro}, Hunsaker and co-workers \citep{reid2021general,goates2023modern} introduced refinements to extend LLT to swept wings. In {Fig.}~\ref{fig:SpanwiseCLb}, we consider sweep $=30^\circ$ and  {AR=133.33}. LLT, PANAIR, and LLT+NN agree over most of the span; however, LLT slightly overpredicts near the root and does not capture the root lift loss at $y=0$, a known effect \citep{kuchemann1953distribution}. LLT+NN recovers this behavior, indicating that the network learns salient physics absent from LLT. 

{Figs.}~\ref{fig:SpanwiseCLa} and \ref{fig:SpanwiseCLb} are in-distribution samples (geometric and aerodynamic parameters within the training range). {Fig}~\ref{fig:SpanwiseCLc} extrapolates in AR. {Figs.}~\ref{fig:SpanwiseCLd}--\ref{fig:SpanwiseCLf} are further out-of-distribution (OOD) cases: taper ratio $=0.1$, sweep $=45^\circ$, and twist $=7.5^\circ$ at AR $=4.0$, respectively. In all OOD cases, LLT+NN remains close to PANAIR, suggesting that training on PANAIR while conditioning on LLT inputs enables the network to learn generalizable trends. 

Spanwise induced-drag distributions for LLT, PANAIR, and LLT+NN are shown in {Fig.}~\ref{fig:SpanwiseCD}. In all cases, LLT deviates noticeably from PANAIR, whereas LLT+NN closely follows the high-fidelity reference.

{Although the neural network itself is not mechanistically interpretable, the corrections to LLT are phenomenologically structured. Defining
	$\Delta c_l(y)=c_l^{\mathrm{LLT+NN}}(y)-c_l^{\mathrm{LLT}}(y)$ and
	$\Delta c_d(y)=c_d^{\mathrm{LLT+NN}}(y)-c_d^{\mathrm{LLT}}(y)$, we observe that the amplitude of $\Delta c_l$ decreases with increasing aspect ratio and becomes negligible at high AR, consistent with LLT’s asymptotic validity (e.g., Fig.~\ref{fig:SpanwiseCLa}). For backward swept wings, the residual concentrates near the root and recovers the centerline unloading captured by lifting-surface/panel methods but absent from classical LLT (Figs.~\ref{fig:SpanwiseCLb}, \ref{fig:SpanwiseCLe}, \ref{fig:SpanwiseCLf}). Together, these patterns indicate that the grey-box model enriches LLT with higher-order three-dimensional effects rather than fitting case-specific noise.}
	
\begin{figure*}[ht]
	\centering
	\subfloat{\includegraphics[width=0.48\textwidth]{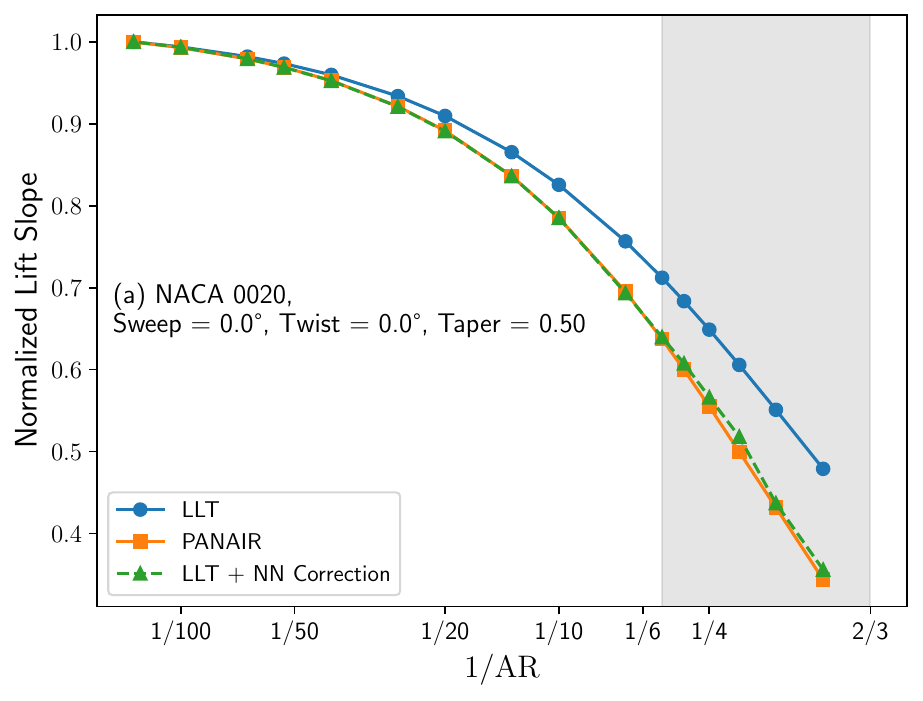}\label{fig:LiftSlopeMostlyIndista}}
	\hfill
	\subfloat{\includegraphics[width=0.48\textwidth]{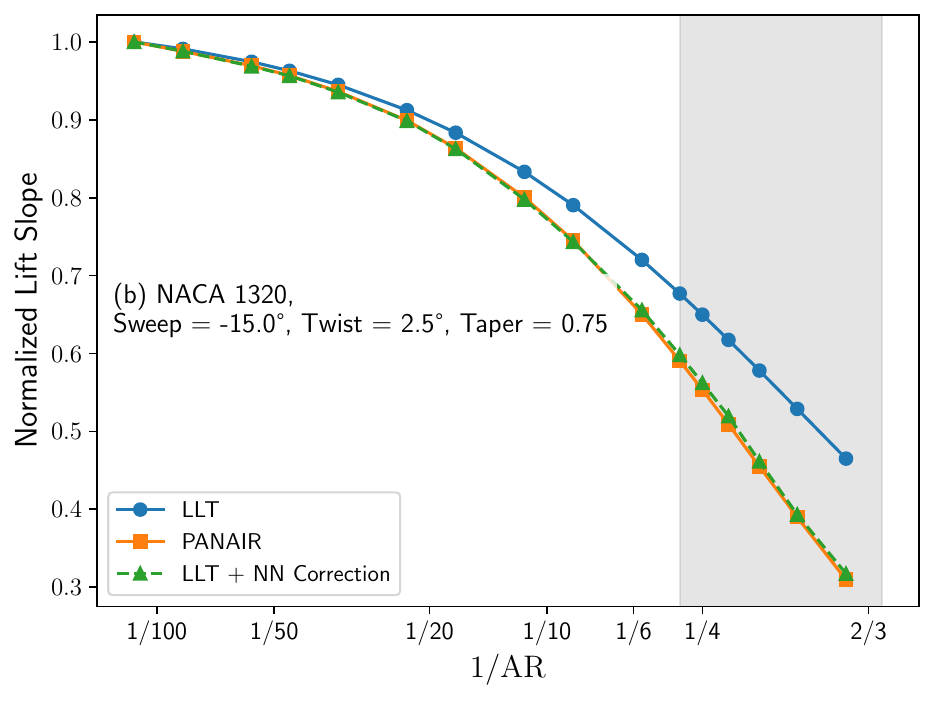}\label{fig:LiftSlopeMostlyIndistb}}
	\vspace{0.5em}	
	\subfloat{\includegraphics[width=0.48\textwidth]{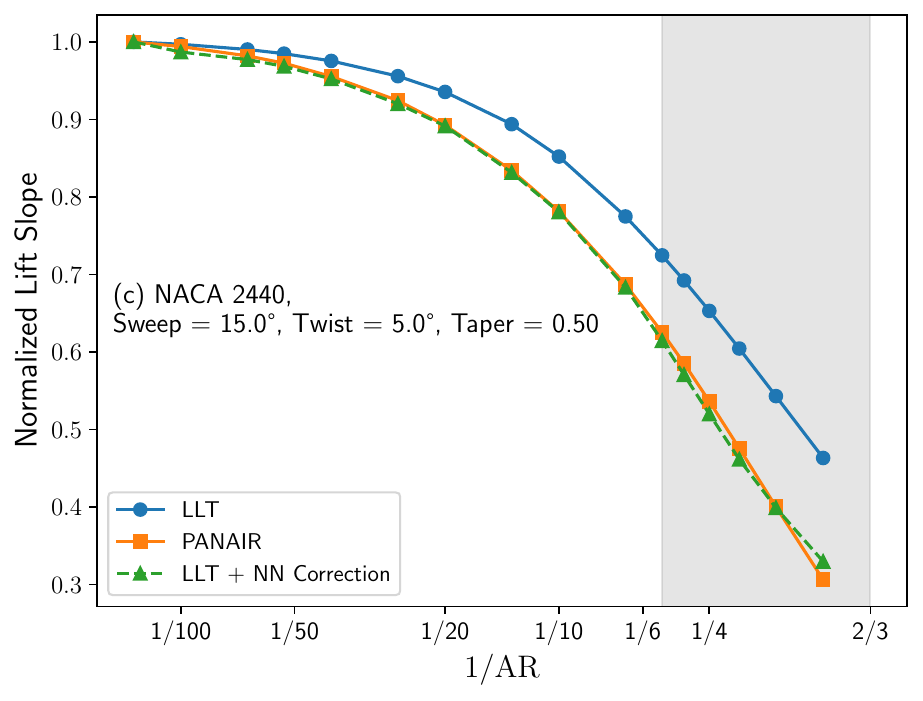}\label{fig:LiftSlopeMostlyIndistc}}
	\hfill
	\subfloat{\includegraphics[width=0.48\textwidth]{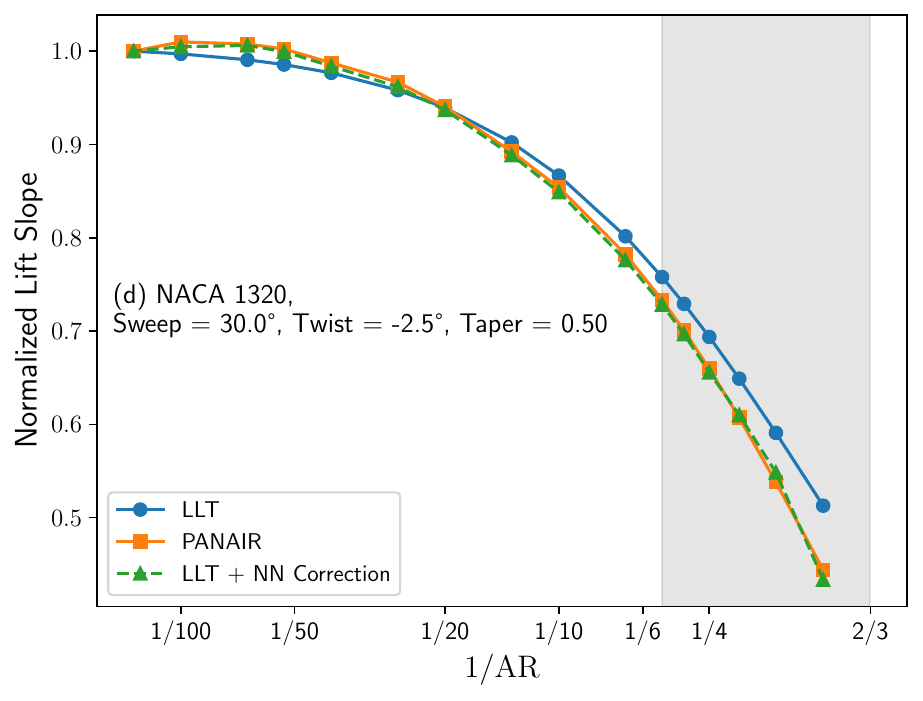}\label{fig:LiftSlopeMostlyIndistd}}
	\vspace{0.5em}	
	\subfloat{\includegraphics[width=0.48\textwidth]{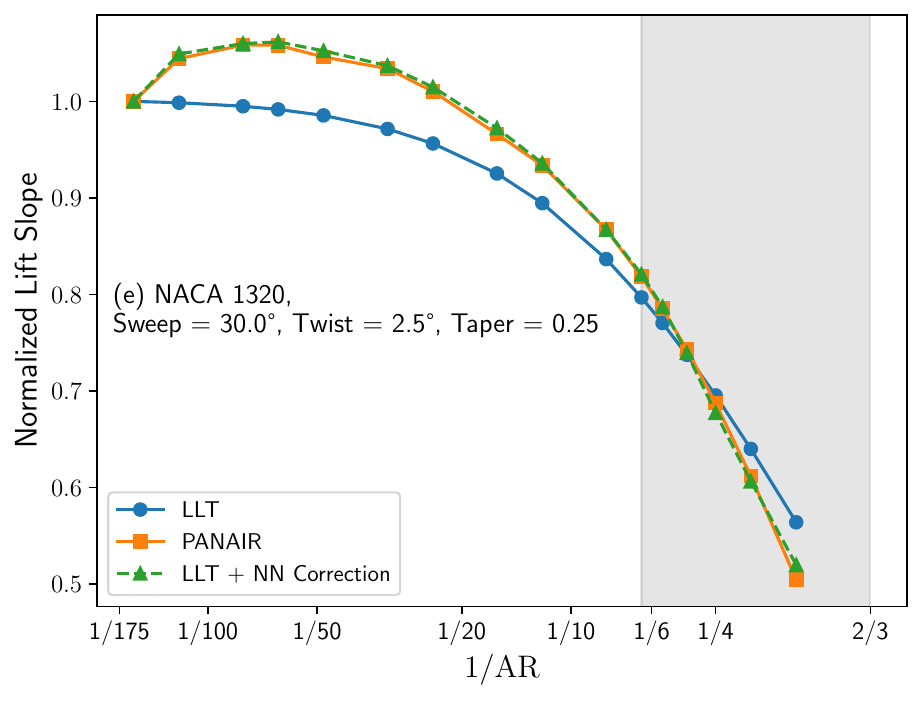}\label{fig:LiftSlopeMostlyIndiste}}
	\hfill
	\subfloat{\includegraphics[width=0.48\textwidth]{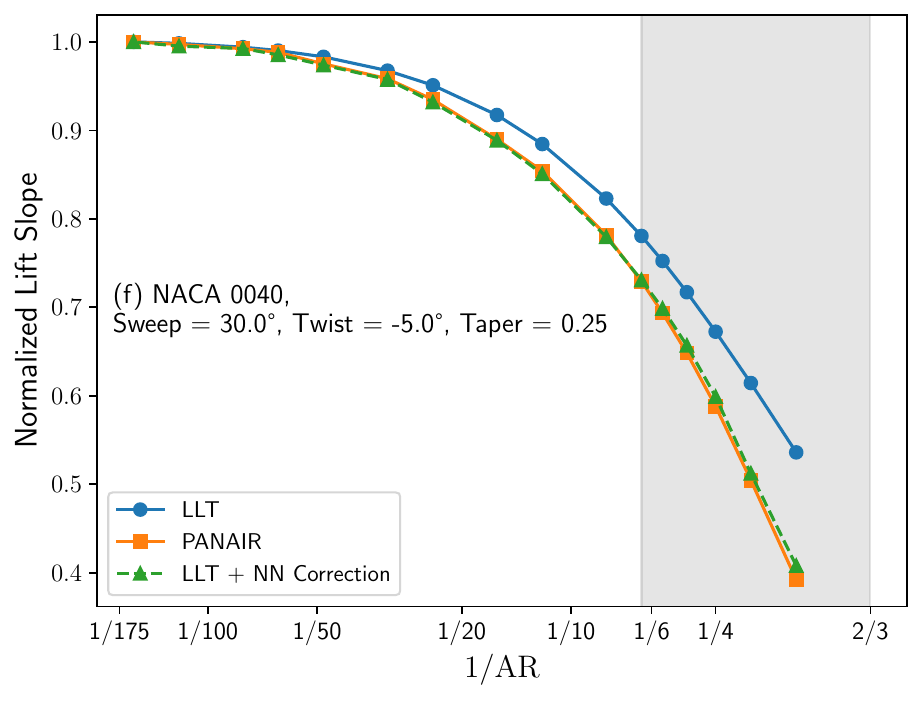}\label{fig:LiftSlopeMostlyIndistf}}
	\caption{Lift-curve slope, normalized by {AR$(1+\lambda)/2=b/c_{\text{root}}$=100}, versus 1/AR with sweep, twist, and taper-ratio ($\lambda$) values within the training-data range. Gray shading marks OOD region {(AR(1+$\lambda$)/2<4 or $b/c_{\text{root}}<4$)}.}
	\label{fig:MostlyIndist}
\end{figure*}

\begin{figure*}[ht]
\centering
\subfloat{\includegraphics[width=0.48\textwidth]{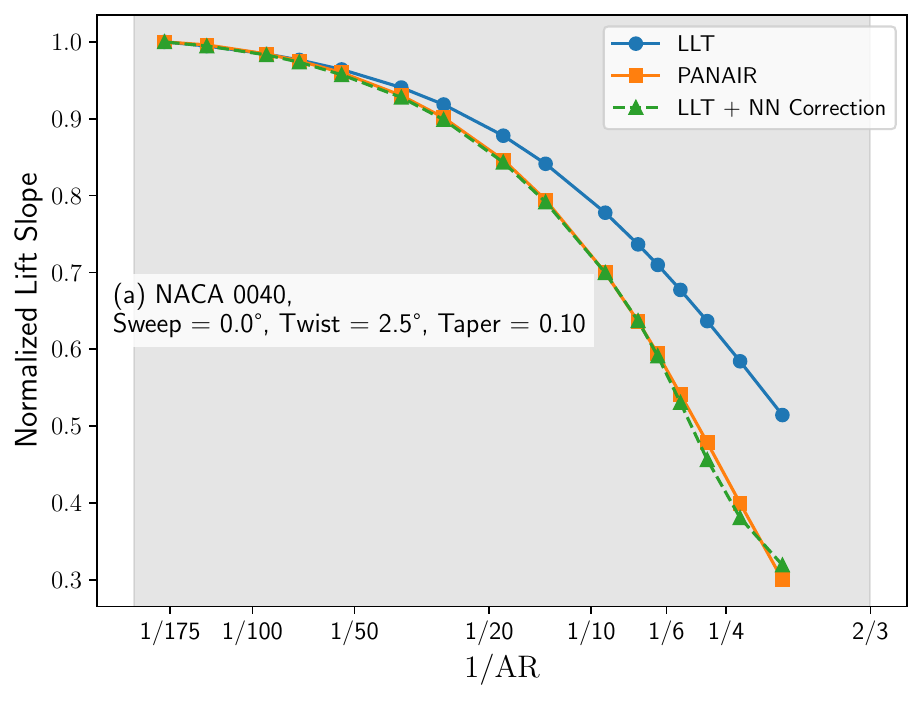}}
\hfill
\subfloat{\includegraphics[width=0.48\textwidth]{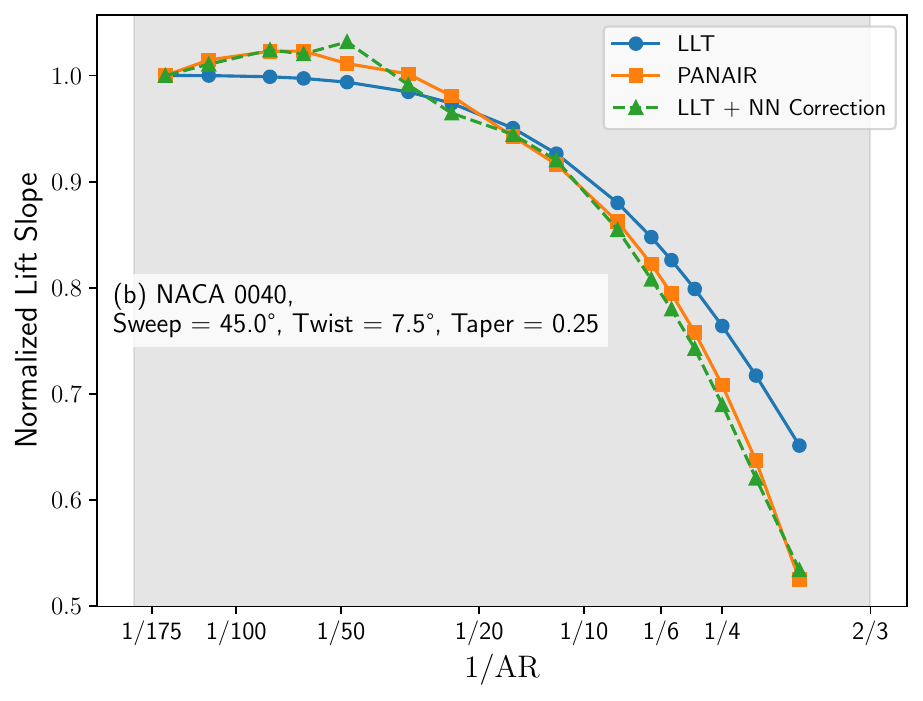}}
\vspace{0.5em}	
\subfloat{\includegraphics[width=0.48\textwidth]{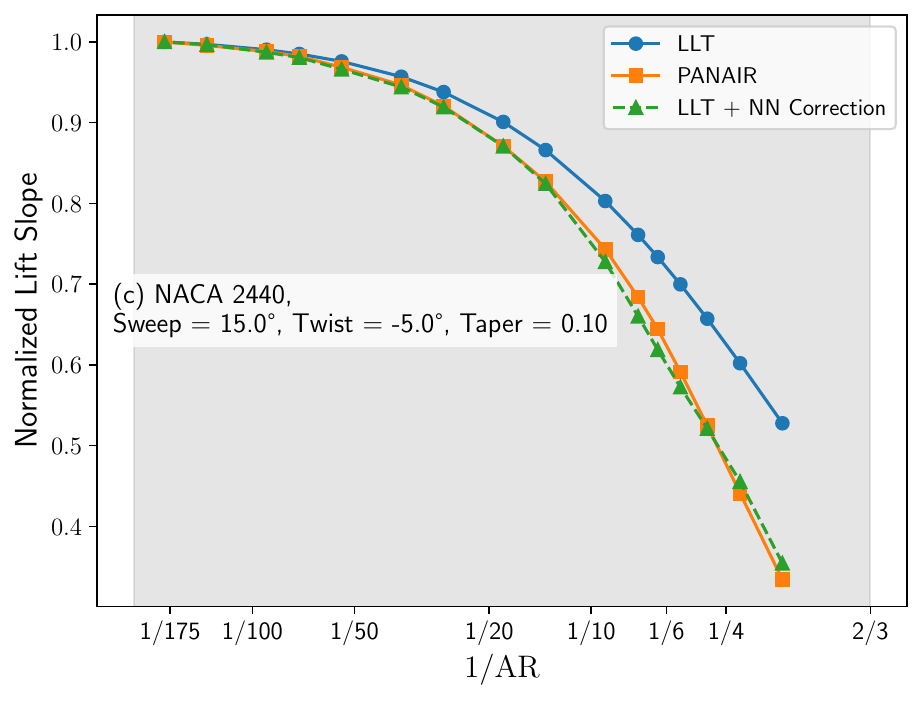}}
\hfill
\subfloat{\includegraphics[width=0.48\textwidth]{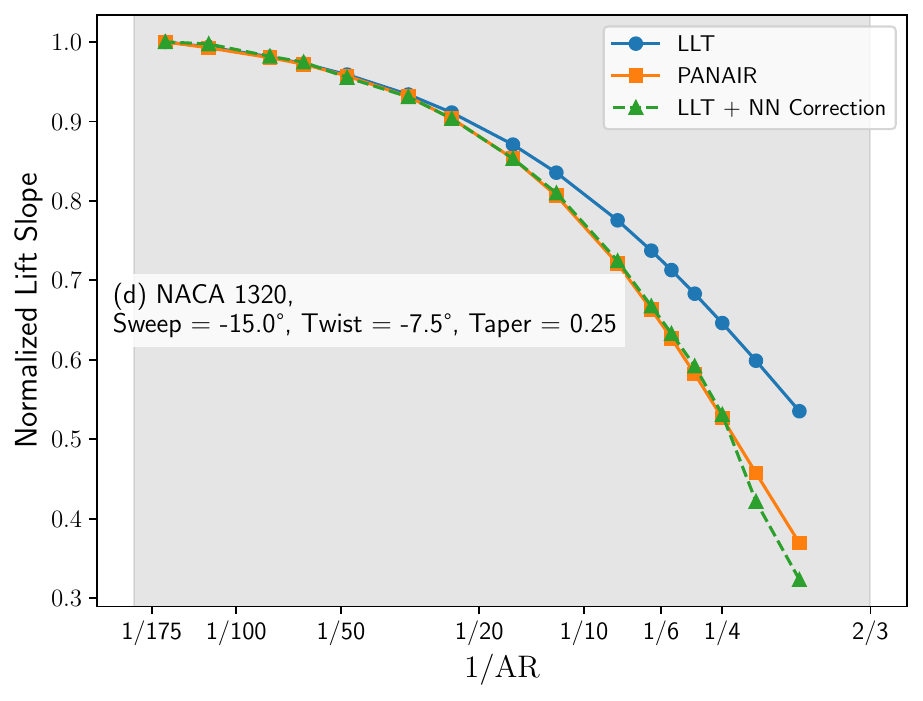}}
\vspace{0.5em}	
\subfloat{\includegraphics[width=0.48\textwidth]{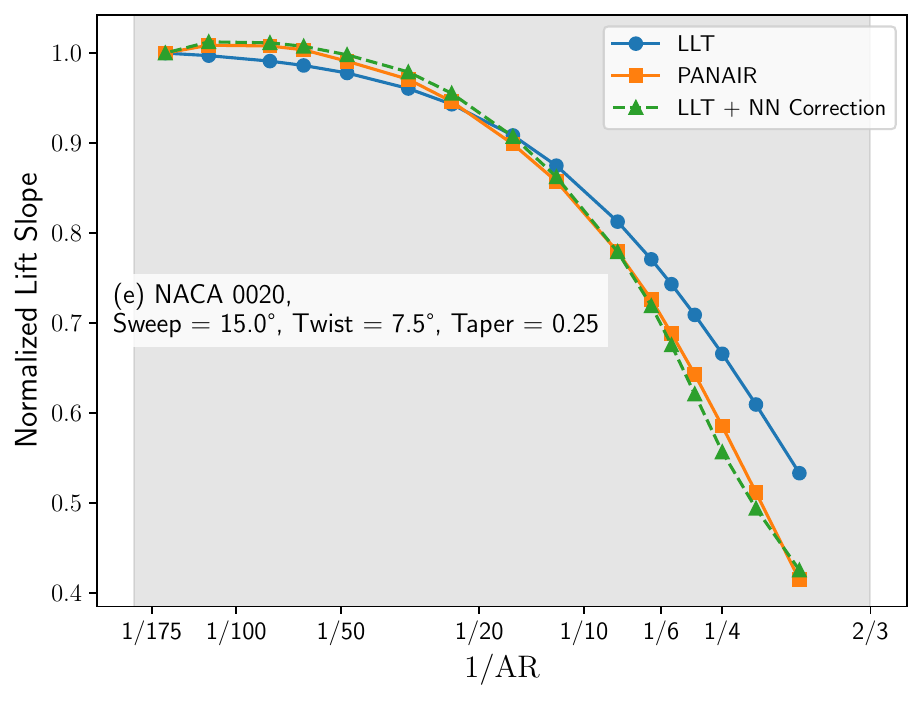}}
\hfill
\subfloat{\includegraphics[width=0.48\textwidth]{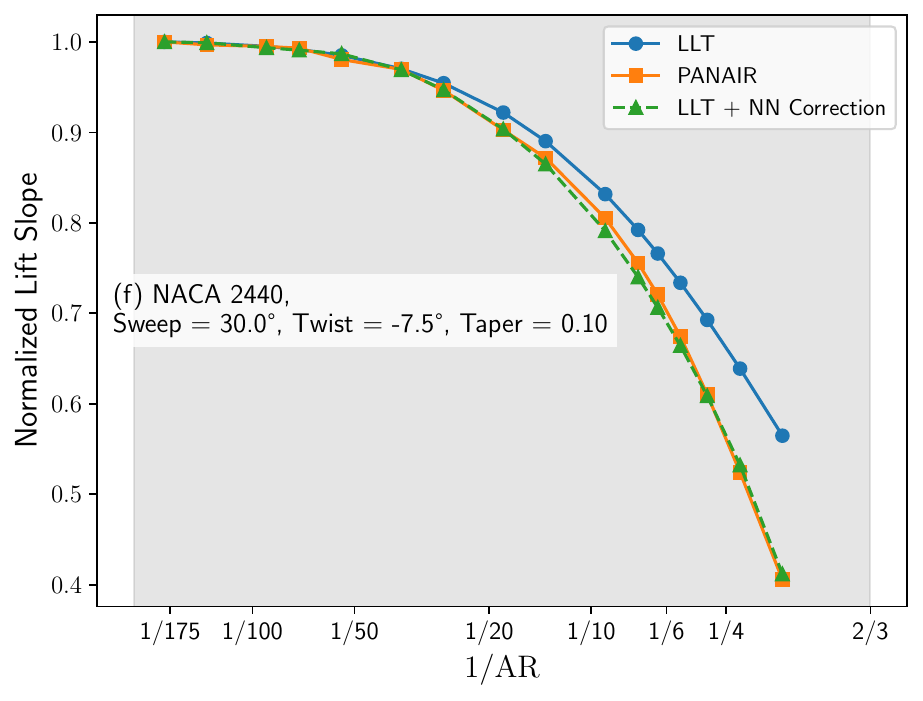}}
\caption{Lift-curve slope, normalized by {AR$(1+\lambda)/2=b/c_{\text{root}}$=100}, versus 1/(AR) for wing shapes with sweep/twist/taper ratio outside the training-data range. Entire domain is OOD and shaded grey.}
\label{fig:Outdist}
\end{figure*}

An important aerodynamic characteristic is the lift-curve slope, the rate of change of lift coefficient with angle of attack. For three-dimensional, unswept wings, this slope is lower than the corresponding two-dimensional airfoil value because of induced effects \citep{anderson2017fundamentals}. {Figs.}~\ref{fig:MostlyIndist} and \ref{fig:Outdist} present lift-curve slope, normalized by its value at {AR$(1+\lambda)/2=b/c_\text{root}$=100}, versus inverse aspect ratio (1/AR). The grey-shaded regions denote wing configurations that lie partially or entirely outside the training/validation ranges. This normalization facilitates a clear comparison of asymptotic convergence behavior as $\mathrm{AR} \to \infty$, where LLT is formally valid. Lift-curve slopes were computed via linear regression over angles of attack from $-6^\circ$ to $6^\circ$ in $0.05^\circ$ increments. For unswept wings ({Fig.}~\ref{fig:LiftSlopeMostlyIndista}), LLT and PANAIR converge as $1/\mathrm{AR} \to 0$, with increasing discrepancy at lower AR. Similar trends hold for sweep $\pm15^\circ$ ({Figs.}~\ref{fig:LiftSlopeMostlyIndistb} and \ref{fig:LiftSlopeMostlyIndistc}) and for a $30^\circ$ swept wing with a symmetric NACA~0040 airfoil ({Fig.}~\ref{fig:LiftSlopeMostlyIndistf}). Across all cases, LLT+NN closely follows PANAIR over the full AR range, including low values near $\mathrm{AR}=1.5$, indicating accurate corrections beyond LLT’s nominal regime.

{Neural networks have been observed to internalize governing constraints; for example, ~\cite{peng2022learning} showed that a network trained to predict the sectional lift coefficient from airfoil geometry and angle of attack can implicitly learn the Kutta condition.} Earlier efforts by \cite{guermond1990generalized} and \cite{kida1978alternative} extended LLT to lower aspect ratios using asymptotic expansions in inverse aspect ratio. {It is possible that, in the present study, by analogy with \cite{peng2022learning}}, LLT+NN has implicitly learned asymptotic corrections by using PANAIR data. This is supported by the good agreement between LLT+NN and PANAIR in the grey-shaded regions of the plots, which indicate AR values outside the training domain.

As noted in the discussion of {Fig.}~\ref{fig:SpanwiseCLb}, LLT does not capture the root lift drop for a $30^\circ$ swept wing at {AR=133.33}. Consistent deficiencies lead to inaccurate lift-curve slopes across the AR range in {Figs.}~\ref{fig:LiftSlopeMostlyIndistd} and \ref{fig:LiftSlopeMostlyIndiste} (sweep $=30^\circ$, NACA~1320). LLT+NN reproduces PANAIR values in these cases as well. While {Fig.}~\ref{fig:MostlyIndist} shows in-distribution samples, {Fig.}~\ref{fig:Outdist} presents OOD wing shapes; even for these challenging cases, LLT+NN remains close to PANAIR.


\section{Conclusions and future work}\label{sec:Conclusions}
We demonstrated a physics-informed neural network that integrates the structure of classical lifting-line theory (LLT) with high-fidelity aerodynamic data from the PANAIR panel method. Using LLT-consistent inputs and outputs, the trained model substantially improves accuracy in predicting spanwise lift and drag distributions for finite wings, including scenarios outside the original training dataset. This behavior suggests that the network captures generalizable aerodynamic trends, implicitly learning higher-order effects often described by asymptotic expansions in inverse aspect ratio.

{The robust predictive capability across varied wing geometries, including sweep, twist, taper, and aspect ratio variations, positions the model as a reliable aerodynamic analysis tool.} Importantly, it addresses traditional LLT limitations, such as inaccuracies at low aspect ratios and significant wing sweep, thus bridging computational efficiency and accuracy. The model shows promise for preliminary optimization tasks, including optimizing wing chord and twist distributions for tractor-propeller aircraft \cite{rakshith2015optimal,sharma2024wing} and analyzing twist effectiveness of control surfaces \cite{montgomery2025lifting}.

Our findings suggest that predictions derived from detailed three-dimensional discretizations (such as PANAIR) can effectively enhance simpler one-dimensional lifting-line approximations. Future work could leverage more accurate datasets, such as computational fluid dynamics (CFD) simulations or wind tunnel experimental data, to further refine model accuracy. Given that such datasets will be typically smaller, approaches could include either data fusion with PANAIR samples with weighted accuracy or adopting a foundation model frameworks seen in weather forecasting \cite{bodnar2025foundation} and biomolecular modeling \cite{abramson2024accurate} applications. Following the foundation model approach the current model, pre-trained on extensive PANAIR data, could serve as a first phase of training and the limited high-fidelity datasets could be utilized to fine tune the model for specific aerodynamic shapes. 

The effectiveness of the neural network notably depends on LLT, as demonstrated by the superior performance of the grey-box approach over purely black-box models. Similar methodologies could extend to other aerodynamic prediction scenarios. For example, future studies might utilize data from blade-resolved or virtual blade element methods \cite{wahono2013development}, which account for turbulence, to enhance rotor lifting-line theories \cite{epps2013unified,falissard2024use}, pushing beyond moderate load and chord variation limits. The present model is trained on inviscid, incompressible panel-method data and does not model viscous separation or compressibility; extending the dataset and hybridizing with viscous/transonic corrections are natural next steps.

\section*{Funding Sources}
{This work was supported by the U.S. Department of Energy, Office of Science, Advanced Scientific Computing Research (ASCR) Early Career Research Program. This paper describes objective technical results and analysis. Any subjective views or opinions that might be expressed in the paper do not necessarily represent the views of the U.S. Department of Energy or the United States Government. Sandia National Laboratories is a multimission laboratory managed and operated by National Technology and Engineering Solutions of Sandia, LLC, a wholly owned subsidiary of Honeywell International Inc., for the U.S. Department of Energy’s National Nuclear Security Administration under contract DE-NA0003525.}

\section*{{Acknowledgments}}
{We are grateful to the anonymous reviewers for their valuable suggestions.}

\bibliography{DataDrivenLLT}

\end{document}